# Simulation Methodology for Analysis of Substrate Noise Impact on Analog / RF Circuits Including Interconnect Resistance

C. Soens[(1,2)], G. Van der Plas[(1)], P. Wambacq[(1,3)], S. Donnay[(1)]
(1) IMEC – (2) also Ph.D. student at ETRO-ELEC, Vrije Universiteit Brussel, Belgium
(3) also lecturer at Vrije Universiteit Brussel, Belgium

**Abstract**

*This paper reports a novel simulation methodology for analysis and prediction of substrate noise impact on analog / RF circuits taking into account the role of the parasitic resistance of the on-chip interconnect in the impact mechanism. This methodology allows investigation of the role of the separate devices (also parasitic devices) in the analog / RF circuit in the overall impact. This way is revealed which devices have to be taken care of (shielding, topology change) to protect the circuit against substrate noise. The developed methodology is used to analyze impact of substrate noise on a 3 GHz LC-tank Voltage Controlled Oscillator (VCO) designed in a high-ohmic 0.18 µm 1PM6 CMOS technology. For this VCO (in the investigated frequency range from DC to 15 MHz) impact is mainly caused by resistive coupling of noise from the substrate to the non-ideal on-chip ground interconnect, resulting in analog ground bounce and frequency modulation. Hence, the presented test-case reveals the important role of the on-chip interconnect in the phenomenon of substrate noise impact.*

## 1. Introduction

The continuous downscaling of CMOS technology has stimulated the trend to integrate RF circuits and digital base-band processing circuits onto one single chip. This has enabled a cost reduction and therefore a proliferation of mixed-mode designs during the past years. However, this single chip solution is facing the problem of analog performance degradation due to impact of digital switching noise, also called substrate noise.

The problem of substrate noise coupling is in general divided into three main fields of study [1]: generation, propagation and impact of substrate noise. Actual simulation methods contain an RC-model of the substrate and a package model [2,3,4]. These are used to simulate substrate noise coupling from a digital circuit to any location on the substrate (e.g. under a sensitive analog circuit). These simulation methods are extended with macro-modeling techniques to efficiently simulate the substrate noise generation of large digital circuits [5,6]. For the analysis of substrate noise impact on the sensitive analog / RF circuits up to now no satisfying simulation method exists. The substrate noise simulation methodology presented in [3] for instance, only models the substrate noise waveforms up to the sensitive nodes at the interface of the substrate and the analog / RF circuit. A comprehensive simulation approach taking all the important effects into account to model the impact of these waveforms remains elusive. Interconnect, for instance, is so far studied as a source of substrate noise [7] (capacitive coupling of noise from noise source interconnect into substrate) or as a path for substrate noise coupling (capacitive coupling of noise from substrate into interconnect of analog victim) [8]. However, the role in the impact mechanism (coupling of noise from substrate into interconnect of the analog circuit resulting in a voltage drop in the circuit) is not yet taken into account. This work reports a simulation methodology for analysis and prediction of the impact of substrate noise on an analog / RF circuit. The method takes interconnect parasitics into account in addition to all previously known effects. The waveforms, resulting from substrate noise impact, can be predicted by simulation for all the nodes at the interface between the substrate and the circuit as well as within the circuit (including the circuit output node). This allows detailed analysis of the impact mechanism. The presented method is validated by comparison of simulation results with measurements on an LC-tank VCO. A simulation accuracy of 2 dB is obtained. The analysis of the test chip reveals that the on-chip ground wire parasitics have an important influence on the level of substrate noise impact. Based on this, a model for impact of substrate noise on an LC-tank VCO was developed [9]. The presented impact simulation methodology when combined with a generation modeling methodology [10] would permit mixed-signal chip verification and sign-off of substrate noise coupling issues.

Section 2 briefly describes the phenomenon of substrate noise impact on an analog / RF circuit. In Section 3 the impact simulation methodology is explained and validated by applying it to a simple one-transistor circuit. In section 4 the VCO used for further validation of the simulation methodology is described. Section 5 explains the mechanisms behind coupling and impact of substrate noise on an LC-tank VCO. Section 6 resumes the experimental and simulation results on the 0.18 µm VCO. Finally conclusions are drawn in section 7.

## 2. Substrate noise coupling and impact

Figure 1 illustrates the problem of the impact of substrate noise on an analog / RF circuit as well as the parasitic contributions influencing this impact. A substrate noise signal generated at a distant location from its analog victim will first travel trough the substrate (**Figure 1** (**a**)) up to the different sensitive nodes at the interface between the substrate and the circuit (**Figure 1** (b)), for instance an NMOS back-gate. The substrate signal will then couple into the circuit via this back-gate (**Figure 1** (I)) and propagate through the circuit to finally impact its output. It is thus necessary to extract a good model for the substrate, the circuit but as well the interface between both. In addition, supply network parasitics (**Figure 1** (**c**)) can influence the substrate noise coupling and impact. Last but not least, the parasitics of the on-chip interconnect play an important role.

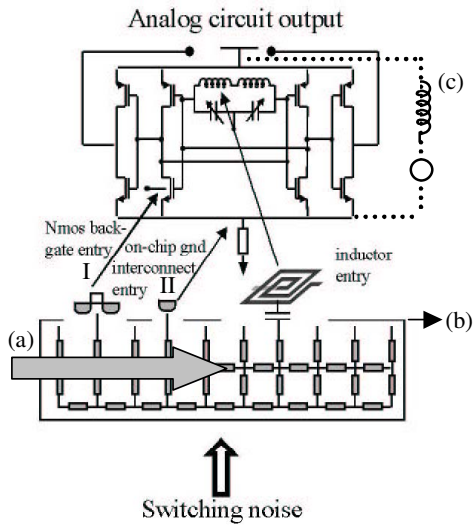

**Figure 1 The presented simulation method allows the analysis of impact of substrate noise through the different sensitive components in the analog / RF circuit. For a VCO important entries into the circuit are the interconnect, the NMOS back-gate and the inductors.**

A substrate noise signal can couple resistively (via e.g. substrate contacts connecting the on-chip metal ground lines with the substrate, **Figure 1** (II)) or capacitively to the circuit interconnect causing a voltage drop in this interconnect and resulting in a voltage variation in the circuit. This voltage variation will then further propagate through the circuit to its output node. This 'interconnect related impact' can seriously harm the analog / RF performance. Hence a good model of the on-chip interconnect is mandatory for prediction of substrate noise impact. Therefore, unlike the classical simulation methodologies [2,3,4,7,8], the proposed methodology (**Figure 2**) employs a model of the interconnect parasitics for simulation of the impact of substrate noise on analog / RF circuits.

## 3. Impact simulation methodology

Analysis starts from the layout of the analog / RF circuit (**Figure 2**). Process technology information is fed to the substrate modeler, the circuit extractor and the interconnect modeler to extract an RC-model for the substrate, an analog / RF circuit model and an interconnect model respectively. To these models a package model is added to obtain a simulation model for the entire system. This complete model is fed to the impact simulator together with the substrate noise waveforms. Finally, waveforms resulting from substrate noise impact can be predicted for the nodes at the interface between the substrate and the analog / RF circuit, as well as all the nodes within the circuit, including the circuit output node.

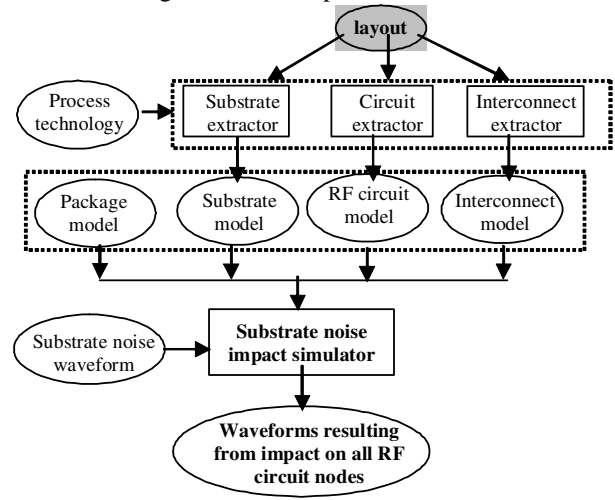

**Figure 2 Substrate noise impact simulation methodology.**

For more detailed substrate noise impact investigations, impact through the different sensitive (parasitic) devices of the circuit (important devices for a VCO are illustrated in **Figure 1**) can be analyzed separately. It provides an analog designer with crucial information on which devices in the circuit have to be carefully shielded or which topology changes can be made to increase the substrate noise immunity of the circuit.

As an example, this paragraph will compare measurement and simulation of the impact of a substrate signal on an NMOS. The NMOS back-gate is the node in an analog circuit that is commonly suggested as the path via which substrate noise is likely to enter the circuit. In **Figure 3** measurements of the transfer from the substrate to the NMOS output are compared to simulations. The coupling from the substrate to the circuit via the NMOS back-gate is modeled taking into account the influence of the interconnect resistance. The resistance from the ground ring of the NMOS to the off-chip ground increases the voltage division from the substrate noise source to the back-gate voltage $v_{bs}$ by almost a factor two. This model is sufficient to explain the impact on the circuit within a maximal error of 1 dB.

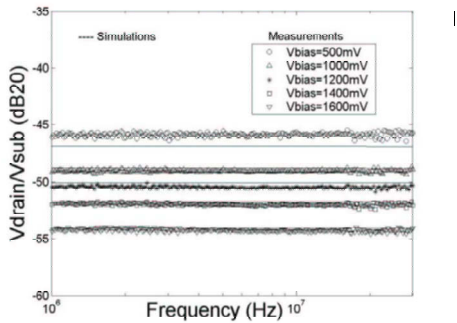

**Figure 3 Measured transfer of a sinusoidal signal in the substrate to the NMOS output agrees well with simulations. This reveals that the NMOS back-gate in combination with the parasitic resistance in the ground interconnect is the dominant path via which substrate noise impacts the NMOS.**

The following paragraph will describe more in detail the models for the substrate, the NMOS and the interconnect extracted with Substratestorm [11] and DIVA [12] and simulated with Spectre RF [13]. **Figure 4** (a) shows the layout of the studied RF NMOS surrounded by a ring of contacts (here called NMOS Ground Ring (*MOS GR*)) connecting the substrate to the ground. **Figure 4** (b) shows the layout of the NMOS measurement structure. Four NMOS transistors are connected in parallel to realize a high enough back-gate transconductance ($g_{mb}$) for sufficient measurement accuracy. **Figure 4** (c) shows a simplified view of the NMOS measurement structure with the locations on the substrate of the four important 'nodes' of the problem: the contact via which a signal is injected into the substrate (*SUB*) the NMOS back-gates (*Back-gate*), the NMOS ground ring (*MOS GR*) and the ground ring surrounding the complete measurement structure (*GR*). **Figure 4** shows the extracted macro-model of the substrate and interconnect (d), connected to the back-gate of the RF NMOS model (e). From the macro-model of the substrate the voltage division from the substrate contact (*SUB*) to the voltage at the NMOS back-gate ($v_{bs}$) equals 1/652. The transfer of noise from substrate to NMOS output (measured and simulated in **Figure 3**), can be computed by multiplying this factor with the back-gate transconductance ($g_{mb}$) and the output impedance of the NMOS ($g_{ds}$). $G_{mb}$ equals 10 mS to 38 mS and $g_{ds}$ 2.8 mS to 22 mS for a bias voltage of 500 mV to 1.6 V. Hence the total transfer ($(v_{back\_gate}/v_{sub}) \, g_{mb} \, r_{ds}$), computed from these values, equals approximately -45 dB to -52 dB and is thus very close to the simulated values shown in **Figure 3**. This confirms that the back-gate together with the resistance in the ground interconnect play the dominant role in this impact problem. Moreover, the good agreement obtained between measurement, simulation and 'hand calculation' validates the models of impact via an NMOS back-gate and of the influence of interconnect parasitics, contained in our simulation model. Coupling via the small source-bulk and drain-bulk junction capacitors ($C_{dbj} = 120 \, fF$, $C_{sbj} = 200 \, fF$) is negligible for the studied frequency range. Impact via the source-bulk and drain-bulk junction capacitors becomes comparable to impact via the back-gate at frequencies between 5 GHz and 19 GHz for bias voltages from 0.5 V to 1.6 V ($f_{3dB} = gmb/2\pi(C_{dbj} + C_{xbj})$).

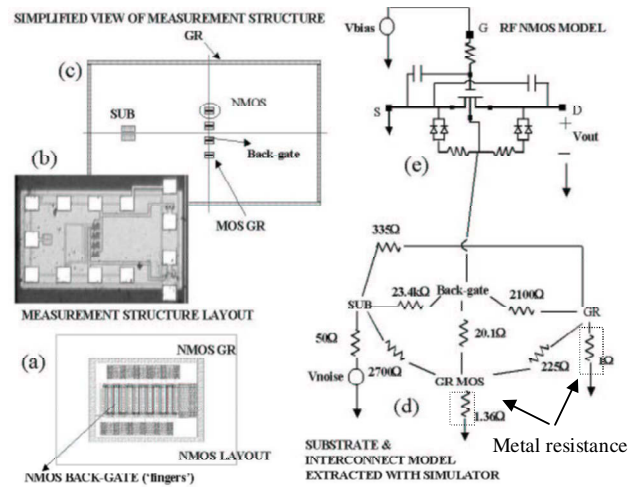

**Figure 4 Overview of (a) the NMOS measurement structure, (b) the RF NMOS layout, (c) a simplified view of the measurement structure, (d) the extracted substrate and interconnect model and (e) the RF NMOS model.**

## 4. Test chip

This section briefly describes the test chip and the experiment used to validate the simulation methodology described in section 3. A VCO designed in a high-ohmic (20 ohm cm) twin-well 1P6M 0.18 µm CMOS technology (schematic shown in **Figure 5** and microphotograph shown in **Figure 6**) is used as a substrate noise victim. It uses an NMOS / PMOS cross-coupled pair and an LC-tank formed by an on-chip inductor and an accumulation mode NMOS varactor. It operates around 3 GHz and has a phase noise of -100 dBc / Hz @ 100 kHz with a current consumption of 5 mA (VCO core) and a supply voltage of 1.8 V.

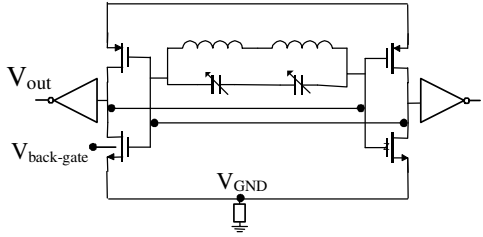

**Figure 5 Schematic of the 0.18 μm CMOS LC-tank VCO: two entries to the circuit for substrate noise are indicated: the on-chip ground ($V_{GND}$) and the NMOS back-gate ($V_{back-gate}$).**

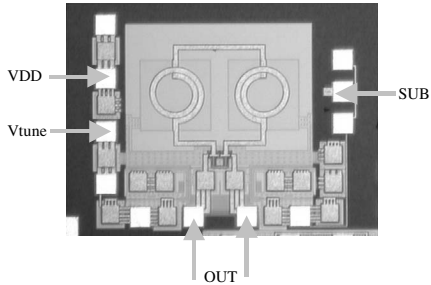

**Figure 6 Microphotograph of the VCO: Vtune pad, $V_{DD}$ pad, output pads, and pad to substrate contact for noise injection (*SUB*).**

A –5 dBm sinusoidal signal is injected into the substrate in the vicinity of the circuit (**Figure 6** *SUB*). The impact of this substrate signal is measured and simulated. The circuit is powered with RF probes and its output is measured single ended with a spectrum analyzer (HP 8565E). The references (ground) for on one hand the sinusoidal signal injected into the substrate and, on the other hand the VCO input and output signals are connected on-chip.

## 5. Substrate noise impact mechanism for an LC-tank VCO

Substrate noise can couple into an LC-tank VCO (LC-tank consisting of an accumulation mode NMOS varactor and an on-chip inductor) through several sensitive components of the circuits:
- resistively and capacitively to the non-ideal metal ground interconnect;
- resistively to the NMOS transistor back-gates;
- capacitively through the n-well to the PMOS transistor back-gates;
- capacitively to the inductors;
- capacitively through the n-well of the accumulation mode NMOS varactor, to its back-gate node;
- capacitively to the on-chip power supply interconnect;

Resistive coupling to the VCO is independent of the noise frequency while capacitive coupling is proportionally increasing with the noise frequency. All these effects are taken into account in the proposed methodology. When a substrate noise signal couples into the VCO, it will modulate the oscillator signal both in amplitude and frequency. The output voltage of the VCO, $V_{out}$, can be expressed as follows (n = number of entries for substrate noise in VCO):

$$v_{out}(t) = A_c \left(1 + \sum_i^n G_{AM}^i h_{sub}^i(t) * v_{noise}(t)\right) \cos\left(2\pi(f_c t + \sum_i^n K_i \int h_{sub}^i(t) * v_{noise}(t) \, dt)\right)$$

with $v_{noise}(t) = A_{noise} \cos(2\pi f_{noise} t)$ (1)

With $h_{sub}^i$ the attenuation by the substrate from the noise source to a sensitive component $i$ in the circuit (NMOS back-gate etc), $A_{noise}$ the noise amplitude, $f_{noise}$ the noise frequency, $A_c$ the local oscillator amplitude, and $f_c$ the local oscillator frequency. Further, $K_i$ (Hz / V) represents the sensitivity of the oscillator frequency to a voltage variation on component $i$, and $G_{AM}^i$ ($V^{-1}$) the AM gain associated to a component $i$. Since the substrate noise power is typically small compared to the local oscillator power, narrow-band frequency modulation (FM) can be assumed. Moreover superposition can be used to calculate the overall impact from the contributions of the separate components in the VCO. At the VCO output spurs appear at both sides of the local oscillator (**Figure 7**) at frequencies $f_c \pm f_{noise}$. After some calculations the FM spur amplitude can be written as:

$$|V_{out}^{FM}(f_c \pm f_{noise})| = |\sum_i^n \frac{A_c H_{sub}^i(f) A_{noise} K_i}{2 f_{noise}}| \quad (2)$$

Since a VCO is often used in front of a limiting circuit (e.g. a switching mixer), FM is the most relevant mechanism.

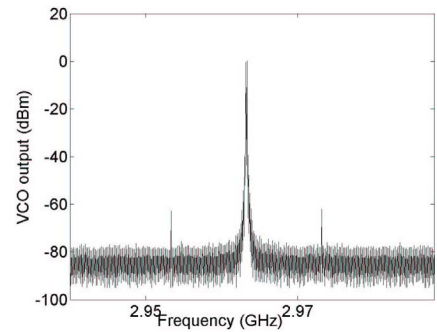

**Figure 7 Power spectrum at the VCO output in presence of a -5 dBm 10 MHz substrate signal: spurs appear at both sides of the local oscillator.**

The amplitude modulation (AM) spurs can be written as:

$$|V_{out}^{AM}(f_c \pm f_{noise})| = |\sum_i^n \frac{A_c H_{sub}^i(f) A_{noise} G_{AM}^i}{2}| \quad (3)$$

From equations (2) and (3) it is clear that in case of resistive coupling ($H_{sub}$ independent of frequency) followed by pure FM the spur voltage is inversely proportional to the noise frequency while for AM it is independent of frequency. These relations will be used in section 6 to investigate the dominant mechanism behind substrate noise impact on the 0.18 μm CMOS VCO.

## 6. Experimental and simulation analysis of substrate noise impact on the VCO

In this section, the simulation methodology described in section 3 is validated with measurements. It is used to investigate the dominant path and mechanism for impact of substrate noise on the 0.18 μm VCO. Models for the substrate, the VCO circuit and the interconnect are extracted with Substratestorm and DIVA. The resulting impact model is simulated using Spectre RF.

As mentioned earlier in this paper, the waveform resulting from impact of substrate noise on the analog / RF victim can be obtained for any node at the interface of substrate and circuit and within the circuit. This allows one to find out what device(s) in the circuit play(s) a dominant role in the overall impact of substrate noise on the circuit performance. It is verified with simulations, that for the studied VCO the non-ideal on-chip ground interconnect (due to the inevitable resistance in this interconnect) is the dominant path for substrate noise impact. Impact via the NMOS transistor back-gate will be shown to be of lesser importance. The impact of a substrate noise signal with a frequency up to 15 MHz is analyzed. Substrate noise signals having these frequencies give rise to spurs close to the local oscillator frequency where disturbing signals are the most harmful. Moreover, the spurs resulting from FM (as will be proven further to be dominant compared to AM) are the largest for this frequency range.

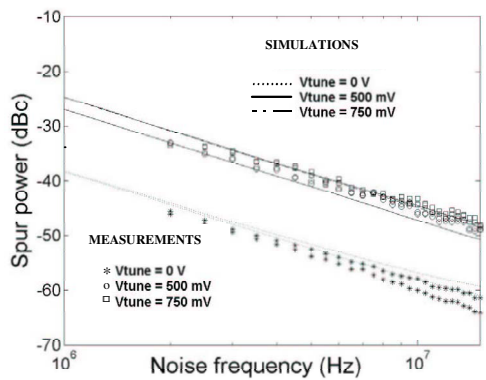

**Figure 8 The total spur power at $f_c \pm f_{noise}$ at the VCO output caused by impact of a -5 dBm sinusoidal substrate signal: a linear relation between the power and the logarithm of $f_{noise}$ is observed.**

Figure 8 shows the comparison between measured and simulated spur power at $f_c \pm f_{noise}$ (for different tuning voltages, *Vtune*) when a –5 dBm sinusoidal signal is injected into the substrate in the vicinity of the VCO (via *SUB* in Figure 6). Simulations of the spur power at left and right side of the local oscillator ($f_c \pm f_{noise}$) agree with measurements with a maximal error of 2 dB, validating the presented methodology. Resistive coupling followed by FM is clearly the dominant mechanism behind impact of substrate noise since the spur power is proportional to the logarithm of the noise frequency. A small difference between left and right spur is observed caused by negligible AM of the local oscillator. In case of resistive coupling followed by AM the spur power would have been independent of the noise frequency. In case of capacitive coupling followed by FM or AM it would have been respectively independent or increasing with frequency. Since capacitive coupling is negligible, it can be concluded that impact via PMOS, inductor and NMOS varactor is negligible as well. The resistive coupling can occur only via the NMOS back-gate or the interconnect parasitic resistance. It will be proven with simulations that the latter is the dominant path via which substrate noise impacts the circuit.

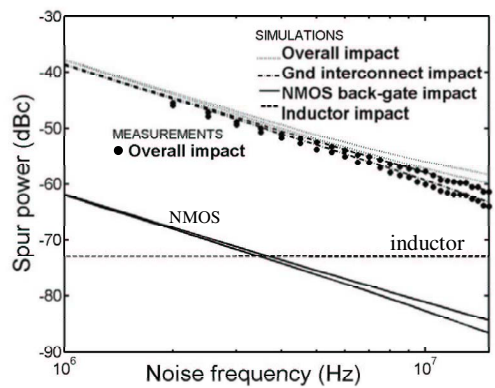

**Figure 9 Simulation of the contributions of the separate devices in the VCO to the overall impact (at $f_c \pm f_{noise}$, Vtune = 0V, Pnoise = - 5dBm). The ground interconnect is the dominant path for substrate noise impact.**

**Figure 9** shows the contribution in the overall impact of the separate components (relevant for the investigated frequency range) in the VCO in terms of spur power in function of noise frequency. The dominant contribution is clearly coming from the parasitic resistance in the on-chip ground interconnect. The inversely proportional relation between spur amplitude and the substrate noise frequency proves that this impact is resulting from a resistive coupling followed by frequency modulation. More precisely substrate noise couples resistively to the ground interconnect, resulting in a voltage drop over its parasitic resistance and causing the voltage over the variable circuit capacitances (NMOS and PMOS capacitances, accumulation mode NMOS varactor capacitance) to vary, which results in modulation of the local oscillator frequency. Impact via the inductor is significantly lower because it results from capacitive coupling ($C_{ind}$ =120 fF per inductor) which is negligible for the studied substrate noise frequencies. It is interesting to note that this impact is constant with frequency proving hat the capacitive coupling is followed by frequency modulation. Impact via the NMOS back-gate is as well inversely proportional to the noise frequency but low compared to the impact via the non-ideal ground interconnect. For our test case a difference of approximately 20 dB, is obtained from simulation in **Figure 9**. Further, impact through the PMOS transistors and the varactor (both in an n-well) is less important than the inductor impact because of the even lower capacitance between the substrate and the n-well they are located in.

Starting from these observations a designer could improve the noise immunity of his circuit by lowering the resistance in the on-chip ground interconnect. A reduction by half of this resistance will lead to a reduction of approximately 6dB in the impact. To test this, a second extraction was done after enlarging where possible the ground interconnect lines in the VCO layout by a factor of two. This yields a prediction of a 4.5 dB lower impact (**Figure 10**).

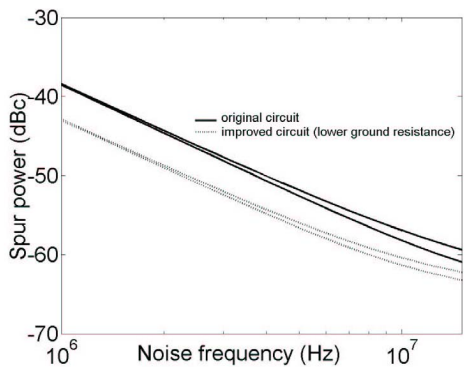

**Figure 10 Simulation of impact (at $f_c \pm f_{noise}$) of a -5 dBm substrate noise signal (1) on the real VCO, (2) a VCO layout with ground interconnect resistance reduced by half.**

For these results simulation requires approximately 35 minutes run time on an HP-UX L2000/4 (4 PA-8600 @ 540MHz) server (extraction time 20 minutes, simulation time 15 minutes for results in figure 10).

## 7. Conclusions

This paper reports a novel simulation methodology for analysis and prediction of substrate noise impact on analog / RF circuits taking into account the parasitic resistance of on-chip interconnect. This methodology allows investigation of the contributions of the separate devices (also parasitic devices) in the analog / RF circuit in the overall impact. This way is revealed which devices have to be taken care of (shielding, topology change) to protect the circuit against substrate noise. The simulation model is validated by comparing measurements to simulations on a 3 GHz VCO designed in a 0.18 μm high-ohmic 1PM6 CMOS technology. Simulations show, that taking the (inevitable) resistance in the on-chip interconnect into account is crucial for accurate prediction of impact of substrate noise on an analog / RF circuit. This methodology has revealed that layout details such as on-chip ground line parasitic resistance can make an analog/RF circuit very sensitive to substrate noise.

## 8. Acknowledgements

The authors wish to tank the IWT Flanders for its financial support and D. Linten for designing the VCO.